# THE LEARNING CRISIS: THREE YEARS AFTER COVID-19

**Tomasz Gajderowicz**
University of Warsaw

**Maciej Jakubowski**
University of Warsaw

**Alec Kennedy**
International Association for the Evaluation of Educational Achievement

**Christian Christrup Kjeldsen**
Aarhus University

**Harry Anthony Patrinos**
University of Arkansas, IZA and GLO

**Rolf Strietholt**
International Association for the Evaluation of Educational Achievement

**Abstract**

The COVID-19 pandemic caused widespread disruptions to education, with school closures affecting over one billion children. These closures, aimed at reducing virus transmission, resulted in significant learning losses, particularly in mathematics and science. Using data from TIMSS 2023, which assesses fourth and eighth-grade achievements across 71 education systems, this study analyzes the impact of school closure duration on learning outcomes. Mixed-effect models estimate deviations from pre-pandemic trends, adjusting for demographic factors. Results show a global average decline of 0.11 standard deviations (SD) in student achievement, with longer closures linked to more severe losses. The effects on low performers, girls, and linguistic minorities show effect sizes up to 0.22 SD. These findings highlight the lasting impact of school closures and emphasize the need for targeted recovery strategies and international cooperation to promote equitable educational outcomes post-pandemic.

Corresponding author:
Harry Anthony Patrinos
University of Arkansas
Fayetteville, AR 72701
USA
E-mail: patrinos@uark.edu

# 1. Background

Coronavirus (COVID-19) is an infectious disease caused by the SARS-CoV-2 virus. Most people infected with the virus will experience mild to moderate respiratory illness and recover without requiring special treatment. However, some will become seriously ill and require medical attention (WHO 2024). Restrictions were established with the intention to reduce the spread of COVID-19, including lockdowns, stay at home orders, and school closures (NBC 2020). One of the arguments for closing schools was that school children at that age might be drivers of the spread within countries and thereby also endangering the older part of the population. However, there is not much evidence that opening schools significantly increased infection rates (Vlachos et al. 2021). During the COVID-19 crisis, one billion children missed one year of schooling, and of these children 700 million missed a total of 1.5 years of education. Many countries were ill-prepared for the extensive school closures, but the few with prior experience of remote learning, trained teachers, appropriate technology, and engaged learners provided learning continuity during the crisis (World Bank 2024). Globally, schools were closed for an average of 5.5 months (22 weeks) since the onset of the pandemic, equivalent to two-thirds of an academic year, when localized school closures are considered (UNESCO 2023). The duration varies by region, from just one month in Oceania, to 2.5 months (10 weeks) in Europe, to as many as 5 months (20 weeks) in Latin America and the Caribbean. Early on, the school closures were expected to contribute to what was already described as a learning crisis (Angrist et al. 2021).

Distance learning during the school closures does not seem to have helped very much, with evidence that it increased dropout risk and lowered test scores (Lichand et al. 2022a; Haelermans et al. 2022; Singh et al. 2022); only the duration of school closures led to variations. Most studies observe increases in inequality where certain demographics of students experienced learning losses that were more significant than others. These learning losses could translate to earnings losses and could cost this generation of students trillions of dollars (Psacharopoulos et al. 2021).

This study explores the global extent of learning loss using TIMSS (Trends in International Mathematics and Science Study) 2023 data, focusing on mathematics and science achievement for grade 4 and grade 8 students. By incorporating school closure durations, we examine how the length and nature of these disruptions affected student outcomes. This is the largest and most comprehensive investigation of learning loss over time, offering a global perspective that includes both pre- and during-pandemic data. We investigate the variation in impacts based on school closure durations and demographic factors, providing a comprehensive view of global trends in educational recovery. Our findings offer valuable insights into the persistence of learning losses in the years following school closures.

We make several contributions to the existing literature. First, we improve the precision of estimates regarding the impact of school closures on learning outcomes. Previous reviews of national studies show large learning losses, with estimates ranging from one-third to half a year's worth of learning (Betthäuser et al. 2023; Donnelly and Patrinos 2021; Hammerstein et al. 2021; Konig and Frey 2022; Panagouli et al. 2021; Patrinos et al. 2023; Sabarwal et al. 2023; Storey and Zhang 2024; Zierer 2021). Reviews of studies produce slightly lower losses than the analyses using the international student assessments. This could be because the reviews have limited samples. For example, Betthauser et al. (2023) review 42 studies and lacks evidence on lower income countries. The international assessments have many more countries: 55 in the case of PIRLS (Jakubowski et al. 2023; Kennedy and Strietholt, 2023) and 71 in the case of PISA (Jakubowski et al. 2024) and are more representative of countries at different income levels. International student assessments, such as TIMSS, offer more comprehensive and representative data across countries with varying income levels, allowing for better cross-country comparisons. The average learning losses identified through large, comparable, international standardized assessments are 0.19 of a standard deviation, or a half years' worth of learning (Crato and Patrinos 2025). TIMSS 2023 offers data up to Spring 2023, providing the most recent measures of student achievement, as prior assessments like PIRLS and PISA were conducted before or during the pandemic (Blanshe and Dahir 2022). Another key contribution of this study is its inclusion of evidence on learning outcomes from both before and after school closures for most countries, with a longer duration of data that covers the entire period of closures for all countries. This allows for a more detailed understanding of the lingering impact of school closures, as prior research has shown long-lasting learning losses (Bollyky et al. 2023; Gambi and De Witte 2024; Hata et al. 2024), although some countries experienced quicker recovery (Miller et al. 2024). The study also highlights the disparities in recovery across countries, with ongoing challenges for many students, particularly those from disadvantaged backgrounds, in regaining lost learning.

Finally, this paper contributes to the growing body of literature on the effects of pandemics and other disruptions to children's schooling (Baker 2013; Belot and Webbink 2010; Ichino and Winter-Ebner 2004; Jaume and Willén 2019; Kóczán 2024). Previous research on past pandemics, such as the 1918 influenza (Ager et al. 2024; Almond 2006) and the 2013–2016 Ebola outbreak (Smith 2021), has shown that school closures result in long-term negative effects on educational attainment and earnings. The 2005 earthquake in northern Pakistan, for instance, demonstrated that children who experienced early life disruptions suffered significant long-term setbacks in academic performance and health (Andrabi et al. 2023). By adding to this literature, our study underscores the lasting impact of school closures on global education, emphasizing the need for targeted recovery efforts to mitigate these losses and address the widening educational inequalities exacerbated by the pandemic.

## 2. Methods

We utilize data from the Trends in International Mathematics and Science Study (TIMSS), an international large-scale, repeated cross-sectional study that randomly samples fourth- and eighth-grade students to assess their achievement in mathematics and science. The curriculum-based assessments measure the knowledge students have accumulated over four and eight years of schooling, respectively. Conducted in four-year cycles, the 2023 cycle marks the first administration following the onset of the COVID-19 pandemic. Drawing on data from six cycles spanning 20 years, starting from 2003, this study examines long-term trends in student performance, allowing us to evaluate whether the 2023 results diverge from pre-pandemic patterns. Altogether, we use data from more than 2.8 million students from 78 countries for Grade 4 and 74 countries for Grade 8. In the most recent TIMSS 2023 cycles the data includes 58 countries for Grade 4 and 44 countries for Grade 8.

TIMSS not only captures student achievement, but also contextual information and characteristics of students through background questionnaires. We use this information as control variables to explain differences in achievement scores while also capturing changes to country demographics or samples over time. Specifically, we use four pieces of information that were consistently captured across all cycles of TIMSS from the student background questionnaire: student age, gender, grade, and how often the language of the test was spoken in the home (*Always or almost always*, *Sometimes*, or *Never*).

Additionally, data on school closure durations are obtained from UNESCO Institute for Statistics (UIS), detailing the number of weeks of full and partial closures in each country. Schools were considered fully closed in cases where government mandates required schools to be closed, affecting most or all students. In contrast, schools were considered partially open when schools were closed only in certain regions or for some grade levels. This also captured schools that were only partially open to in-person instruction (e.g., hybrid learning). This data allows us to assess how learning losses varied by the duration of school closures. For all models, we use a measure of school closures that counts the weeks in which schools were either fully or partially closed.

To capture changes following COVID-19, it is essential to take previous long-term trends into account. Our empirical strategy involves the estimation of deviations of the most recent 2023 result from long-term linear trends in average student mathematics and science achievement for each participating school system (see Jakubowski et al. 2023, 2024; Kennedy and Strietholt 2023). We estimate a separate linear trend for each country and include country-level fixed effects to control for unobserved time-invariant country characteristics. This is formalized in the following regression model:

$$Y_{ijk} = \Sigma_{k=1}^{n} \alpha_k + \Sigma_{k=1}^{n} \beta_k * time + \tau D_{2023} + \gamma X_{ijk} + \varepsilon_{ijk} \tag{1}$$

where $Y_{ijk}$ represents the achievement of student i at school j in country k, with n being the number of countries in the sample being analyzed. The model is estimated on repeated cross-sections, with $\beta_k$ capturing the slope of country-specific linear trends in student achievement across TIMSS cycles. $D_{2023}$ is an indicator variable that is equal to one for the data collected during the 2023 cycle (after the onset of the pandemic) and is zero for all other cycles. $\tau$ is our parameter of interest capturing the deviation from country-specific trends occurring after the onset of the pandemic after controlling for time-invariant country effects ($\alpha_k$) and student background characteristics ($X_{ijk}$). The model is estimated separately for both TIMSS subjects (i.e., mathematics and science) and both student populations (i.e., grade 4 and grade 8).

We next incorporate measures of national school closure policy duration to understand how our estimates of learning loss vary by the length of school closures. This is done through the introduction of an interaction term represented by the regression equation below:

$$Y_{ijk} = \Sigma_{k=1}^{n}\alpha_k + \Sigma_{k=1}^{n}\beta_k * time + \tau D_{2023} + \pi D_{2023} * weeks_k + \gamma X_{ijk} + \varepsilon_{ijk} \qquad (2)$$

with the model specified as it was in Equation 1, with an additional interaction term between $D_{2023}$ and $weeks_k$, the number of weeks that schools were closed in country $k$. Since all countries experienced school closures for some duration, we centered the variable $weeks_k$, to facilitate the interpretation of results, allowing $\tau$ to represent the deviation from the trend for a country with an average length of school closure. $\pi$ then represents how those deviations vary by additional week of school closures for a country.

The model is expanded further to test for differences in the impact of the pandemic and school closure policies by gender and home language. Each variable is dichotomized to provide two different contrasts: girls versus boys, those who report never or only sometimes speaking the test language at home versus those who speak it most of the time. These contrasts are made by adding additional interaction terms between the individual student characteristics and both the $D_{2023}$ and $D_{2023} * weeks_k$ from the model. In addition, the individual characteristics are also interacted with the time trend variable, allowing for trends to be estimated separately by student group.

Finally, we investigate the heterogeneity in the impact of school closures by achievement level. To check how learning losses vary among students of different proficiency, we estimate the main models using quantile regressions, fitting models of different percentiles of TIMSS achievement. To deal with many dummy variables and interactions, we use recent implementations of fast quantile regression algorithms (Chernozhukov et al. 2022). The quantile regression approach has been applied in other international large-scale assessments studies arguing that one need to investigate across the achievement distribution the correlations between background information and outcomes (Kjeldsen & Elezović forthcoming).

Data from TIMSS is collected using a two-stage stratified sampling design with schools drawn as a first stage and one or more classes of students selected from each sampled school in the second stage. To account for this complex sampling design, we used sampling weights (senate weights so each country counts the same to the pooled results) to receive unbiased estimates and present standard errors clustered at the school level. Furthermore, TIMSS uses a rotated booklet design and employs the use of plausible values methodology to estimate students' achievement (Mislevy et al. 1992). Therefore, all results presented are based on estimation accounting for the variation across the five plausible values for each subject scale (Rubin 2004).

## 3. Findings

The results highlight significant learning losses due to school closures, with variations across subjects, grade levels, and demographic factors such as gender and home language. For both Grade 4 and Grade 8 students, performance in mathematics and science fell significantly below expectations based on the linear trends observed in prior TIMSS cycles. Grade 8 students experienced greater declines across both subjects compared to Grade 4 students.

*Learning loss estimates*

Table 1 presents the estimated deviations from long-term trends by subject and grade. Overall, the models indicate significant declines in performance compared to the expected values based on the long-term achievement trends for countries. In Grade 4, the departure from trend was smaller for science than for mathematics. For Grade 8, the deviations from the long-term trends were about the same for both subjects. To put the linear trend departures, we standardize them by the average within-country SD for each grade level and subject ($SD_{Grade4,math}$=84, $SD_{Grade4,science}$=85, $SD_{Grade8,math}$=90, $SD_{Grade4,science}$=91). The effect sizes for Grade 4 were 0.11 SD for mathematics and 0.06 SD for science, while for Grade 8, they were slightly larger at 0.12 SD for mathematics and 0.14 SD for science. These results demonstrate significant negative effects on learning outcomes across all subjects and grade levels.

Table 1: Departure from linear trend

|  | Grade 4 | | Grade 8 | |
| --- | --- | --- | --- | --- |
|  | Mathematics | Science | Mathematics | Science |
| Departure from linear trend | -9.311*** (0.893) | -4.878*** (0.908) | -11.227*** (1.264) | -12.829*** (1.280) |
| N | 1,350,336 | 1,350,336 | 1,473,698 | 1,473,698 |

Standard errors in parentheses. * p<0.05; ** p<0.01; *** p<0.001.

*Learning loss by school closure duration*

Table 2 presents how the deviations from the linear trend vary by duration of national school closures as captured in the UIS database considering both full and partial school closures. Note that four countries had to be omitted from the results due to them not having data collected in the UIS database. In all models, longer school closure durations are associated with greater deviations from long-term trends in TIMSS mathematics and science achievement. That is, countries where schools were closed for longer periods of time fell further below the expected linear trend than countries that closed schools for shorter periods of time.

The duration of school closures was strongly associated with larger deviations from the linear trend, particularly in mathematics. Within grade levels, longer closures were linked to more substantial trend departures in mathematics compared to science. Across grade levels, Grade 8 students experienced more significant declines per week of school closure compared to Grade 4 students, indicating that older students may have been more vulnerable to the extended disruptions in learning.

This negative relationship ranged from about 0.16 points lost per week (Grade 4 Science) up to 0.52 points lost per week (Grade 8 mathematics). This indicates that if schools closed for one school year (approximately 36 weeks), declines would range from 5.62 to 18.68 points, depending on the grade level and subject. This is equivalent to declines in average achievement of between 0.07 to 0.21 SDs.

Table 2: Departure from linear trend and school closure duration

|  | Grade 4 | | Grade 8 | |
|---|---|---|---|---|
|  | Mathematics | Science | Mathematics | Science |
| Departure from linear trend | -9.552*** | -6.266*** | -10.185*** | -13.851*** |
|  | (0.952) | (0.969) | (1.301) | (1.318) |
| School closure duration | -0.247*** | -0.156** | -0.519*** | -0.359*** |
|  | (0.047) | (0.049) | (0.061) | (0.060) |
| N | 1,287,355 | 1,287,355 | 1,409,298 | 1,409,298 |

Standard errors in parentheses. * $p<0.05$; ** $p<0.01$; *** $p<0.001$.

*Learning loss and school closure effects by student background*

The deviations from long-term trends and the variability in school closures differ by gender and home language are reported in Table 3. When examining gender, we find that girls experienced greater deviations from long-term trends than boys across both subjects and grade levels. However, we find no evidence of significant differences in the association between school closure duration and the deviations by gender, suggesting that while girls experienced larger declines in learning after the onset of COVID19 compared to boys, the duration of closures did not disproportionately impact one gender over the other.

The impact of speaking the test language at home was also explored. For Grade 4 students, those who did not speak the test language most often at home experienced greater deviations from the long-term trends than those who did. However, no such difference was found in Grade 8, where trend departures were roughly similar for both groups. Furthermore, there were no significant differences in the association between trend deviations and school closure duration based on home language.

Table 3: Departure from linear trend, school closure duration effects by gender and language

| Estimate | Group | G4 Math | G4 Science | G8 Math | G8 Science |
|---|---|---|---|---|---|
| | | *Heterogeneity by Gender* | | | |
| Departure from linear trend | Girls | -12.510*** | -7.790*** | -16.155*** | -19.871*** |
| | | (1.025) | (1.066) | (1.399) | (1.449) |
| | Boys | -6.641*** | -4.798*** | -4.581** | -8.075*** |
| | | (1.138) | (1.137) | (1.606) | (1.572) |
| | Difference (Girls – Boys) | -5.869*** | -2.992** | -11.574*** | -11.796*** |
| | | (1.040) | (1.067) | (1.548) | (1.544) |
| School closure duration | Girls | -0.243*** | -0.135* | -0.511*** | -0.383*** |
| | | (0.053) | (0.056) | (0.067) | (0.066) |
| | Boys | -0.251*** | -0.179** | -0.531*** | -0.338*** |
| | | (0.059) | (0.062) | (0.076) | (0.074) |
| | Difference (Girls – Boys) | 0.008 | 0.044 | 0.020 | -0.045 |
| | | (0.063) | (0.067) | (0.078) | (0.078) |
| | | *Heterogeneity by home language* | | | |
| Departure from linear trend | Don't speak test language most often | -15.546*** | -12.819*** | -6.197* | -17.651*** |
| | | (1.592) | (1.745) | (2.662) | (2.960) |
| | Speak test language most often | -8.728*** | -5.217*** | -10.707*** | -12.678*** |
| | | (0.949) | (0.958) | (1.277) | (1.249) |
| | Difference (not most often – most often) | -6.818*** | -7.602*** | 4.510 | -4.973 |
| | | (1.524) | (1.682) | (2.484) | (2.706) |
| School closure duration | Don't speak test language most often | -0.226** | -0.157 | -0.371** | -0.393** |
| | | (0.079) | (0.088) | (0.125) | (0.133) |
| | Speak test language most often | -0.251*** | -0.162*** | -0.567*** | -0.373*** |
| | | (0.047) | (0.048) | (0.059) | (0.056) |
| | Difference (not most often – most often) | 0.025 | 0.005 | 0.195 | -0.020 |
| | | (0.077) | (0.084) | (0.115) | (0.123) |
| N | | 1,287,355 | 1,287,355 | 1,409,298 | 1,409,298 |

Standard errors in parentheses. * p<0.05; ** p<0.01; *** p<0.001.

Overall, the findings demonstrate significant deviations from long-term achievement trends in both mathematics and science in the 2023 cycle, with variations by grade level, gender, and home language. These declines were more strongly associated with school closure duration for Grade 8 students compared to Grade 4 students. Additionally, girls faced larger trend departures than boys across both subjects and grade levels. The study also highlights that students who did not speak the test language at home experienced greater deviations from long-term achievement trends in Grade 4, but not in Grade 8. Despite these differences, the duration of school closures did not appear to have a significantly different association with these deviations based on gender or home language. These findings underscore the widespread and varied effects of school closures on student learning.

*Learning loss and school closure effects by achievement percentiles*

Learning loss is notable among low-achieving students but not among those at the top of the achievement distribution. Table 4 presents a comparison of results in mathematics and science for students in grades 4 and 8, focusing on the 10th percentile (low-achievers) and the 90th percentile (high-achievers). The 2023 deviation from the long-term trend is statistically insignificant for high-achieving students. In contrast, low-achieving students across all subjects and grades show a significant decline in 2023 results compared to the country-specific time trends, with learning loss ranging from 0.14 SD to 0.21 SD.

Table 4: Departure from linear trend for low- and high-achieving students

| Estimate | Group | G4 Math | G4 Science | G8 Math | G8 Science |
|---|---|---|---|---|---|
| Departure from linear trend | Low achievers (10th percentile) | -16.469*** (1.258) | -18.024*** (1.677) | -12.665*** (1.265) | -17.473*** (2.371) |
| | High achievers (90th percentile) | -1.164 (2.378) | -1.096 (2.035) | 3.820 (1.977) | -2.928 (2.471) |

Standard errors in parentheses. * $p<0.05$; ** $p<0.01$; *** $p<0.001$.

The results of this analysis highlight significant learning losses due to school closures, with notable variations across subjects, grade levels, and demographic factors such as gender and home language. Both Grade 4 and Grade 8 students showed declines in mathematics and science performance compared to long-term trends, with Grade 8 students experiencing more substantial declines. The duration of school closures was strongly associated with greater learning losses, particularly in mathematics, and older students appeared more vulnerable to prolonged disruptions. Girls generally experienced greater declines than boys, but no significant differences were found in how school closure duration affected learning losses by gender or home language. Grade 8 girls in science suffered the most. Additionally, low-achieving students experience large losses at every grade and subject, while high-achieving students saw no significant deviations. These findings emphasize the broad and varied impact of school closures on student learning outcomes.

## 4. Interpretation

The COVID-19 pandemic caused widespread disruptions in education, leading to lasting impacts on student performance, particularly in mathematics and science. Using TIMSS 2023 data, this study highlights the long-term effects of the pandemic, showing a global decline in student achievement by an average of 0.11 standard deviations compared to pre-pandemic trends. Countries with longer school closures saw more significant declines, with disadvantaged groups experiencing losses of up to 0.22 SD—especially low achievers, girls, and students facing language barriers—suffering disproportionately. The findings emphasize the need for targeted interventions to address these disparities and protect educational equity in the long term.

Moreover, the study's findings emphasize the importance of timely and targeted recovery strategies. To address these learning losses and prevent further deepening of inequalities, the need for international cooperation becomes clear. While the development and distribution of vaccines against the SARS-CoV-2 virus were driven by international cooperation, there has so far been little collaboration to jointly address learning losses and learn from each other. Recovery must involve tailored policies that not only focus on restoring lost content but also on providing additional support to the most vulnerable students. These results further reinforce the call for coordinated global efforts to tackle the educational crisis created by the pandemic and to ensure that future disruptions do not disproportionately affect the most disadvantaged students (Singh et al. 2024).

The findings from this study underscore the need for targeted policy interventions. These include motivational nudges, such as text messages sent to students or their caregivers, which have been shown to significantly increase standardized test scores during remote learning (Angrist et al. 2022). Furthermore, evidence from Australia highlights the importance of targeted funding aimed at promoting more equitable educational outcomes. These efforts, which focused on providing additional resources to the most disadvantaged schools and students, helped mitigate some of the learning loss experienced during the pandemic (Miller et al. 2024). Policy measures that combine direct support for students with efforts to engage parents and communities can be a powerful tool in accelerating recovery and narrowing achievement gaps (Gray-Lobe 2024). During the school closures, high impact, online, tutoring was shown to be cost-effective (Carlana and La Ferrara 2024; Gortazar et al. 2024; Lichand et al. 2022b).

While this study provides valuable insights into the global learning loss caused by school closures, there are several limitations to consider. First, the reliance on TIMSS 2023 data means that the analysis is limited to the countries that participated in the study, excluding nations with lower participation or those that did not assess mathematics and science achievement in the same way. Additionally, while the study adjusts for demographic factors such as gender, socioeconomic status, and home language, it may not fully capture all the complexities of individual student experiences, such as variations in access to online learning resources, teacher quality, or parental involvement during the pandemic. Our study provides average effects at a high (national) level of aggregation; interventions should be targeted, focusing particularly on those who need help the most. The data on school closure durations, while comprehensive, may also be imprecise in some cases, especially in countries with localized or partial closures. In Finland, students who studied on-line for longer periods performed equally well in the matriculation exam at the end of upper-secondary education than the students who experienced shorter school closures (Riudavets-Barcons and Uusitalo 2024). Moreover, school closures may be underestimated. In the case of Sweden, even there were no school closures on a national level, children still became sick, and classes were sent home (Skolverket (2021). Finally, the study focuses on the academic subjects of mathematics and science, which may not fully reflect the broader impact of school closures on other areas of learning, such as language arts or social-emotional development. These limitations suggest that while the findings provide a useful overview of the pandemic's educational impact, further research is needed to explore a wider range of variables and outcomes.